\documentclass[sigconf]{acmart}
\AtBeginDocument{%
  }

\usepackage{wrapfig}
\usepackage{subcaption}
\usepackage{array}

\copyrightyear{2026}
\acmYear{2026}
\setcopyright{cc}
\setcctype{by}
\acmConference[CHI '26]{Proceedings of the 2026 CHI Conference on Human Factors in Computing Systems}{April 13--17, 2026}{Barcelona, Spain}
\acmBooktitle{Proceedings of the 2026 CHI Conference on Human Factors in Computing Systems (CHI '26), April 13--17, 2026, Barcelona, Spain}
\acmPrice{}
\acmDOI{10.1145/3772318.3790662}
\acmISBN{979-8-4007-2278-3/2026/04}

\begin{document}

\title
{SPIRIT: A Design Framework To Support Technology Interventions for Spiritual Care Within and Beyond the Clinic}

\author{C. Estelle Smith}
\affiliation{%
  \institution{Colorado School of Mines, Computer Science}
  \streetaddress{1500 Illinois St, Golden, CO 80401}
  \city{Golden}
  \state{Colorado}
  \country{USA}
  \postcode{80401}}
\email{estellesmith@mines.edu}
\orcid{0000-0002-4981-7105}

\author{Alemitu Bezabih}
\orcid{0000-0001-9603-8537}
\affiliation{%
  \institution{Colorado School of Mines, Computer Science}
\streetaddress{1500 Illinois St, Golden, CO 80401}
  \city{Golden}
  \state{CO}
  \postcode{80401}
  \country{USA}}
\email{alemitubezabih@mines.edu}

\author{Shadi Nourriz}
\orcid{0009-0004-1873-6750}
\affiliation{%
  \institution{Colorado School of Mines, Computer Science}
  \city{Golden}
  \state{CO}
  \country{USA}}
\email{shadinourriz@mines.edu} 

\author{Jesan Ahammed Ovi}
\orcid{0000-0003-0615-621X}
\affiliation{%
  \institution{Colorado School of Mines, Computer Science}
  \city{Golden}
  \state{CO}
  \postcode{80401}
  \country{USA}}
\email{jesanahammed_ovi@mines.edu}

\renewcommand{\shortauthors}{C. Estelle Smith et al.}

\begin{abstract}
Despite its importance for well-being, spiritual care remains under-explored in HCI, while the adoption of technology in clinical spiritual care lags behind other healthcare fields. Prior work derived a definition of "spiritual support" through co-design workshops with stakeholders in online health communities. This paper contributes: (1) a revision of that definition through member checking with professional spiritual care providers (SCPs); (2) a novel design framework---SPIRIT---which can help to expand models of delivery for spiritual care using digital technologies. Through re-analysis of previous data and new interviews with SCPs, we identify three prerequisites for meaningful spiritual care: openness to care, safe space, and the ability to discern and articulate spiritual needs. We also propose six design dimensions: loving presence, meaning-making, appropriate degree of technology use, location, degree of relational closeness, and temporality. We discuss how SPIRIT offers guidance for designing impactful digital spiritual care intervention systems within and beyond clinical settings.
\end{abstract}

\ccsdesc[500]{Human-centered computing~Empirical studies in collaborative and social computing}

\keywords{Spiritual care, spiritual support, social support, spirituality, religion, theology, online health, CaringBridge, prayer, health, patient, caregiver, design}

\newcommand{\ssdefinition}{
\begin{figure}
    \centering
    \includegraphics[width=\linewidth]{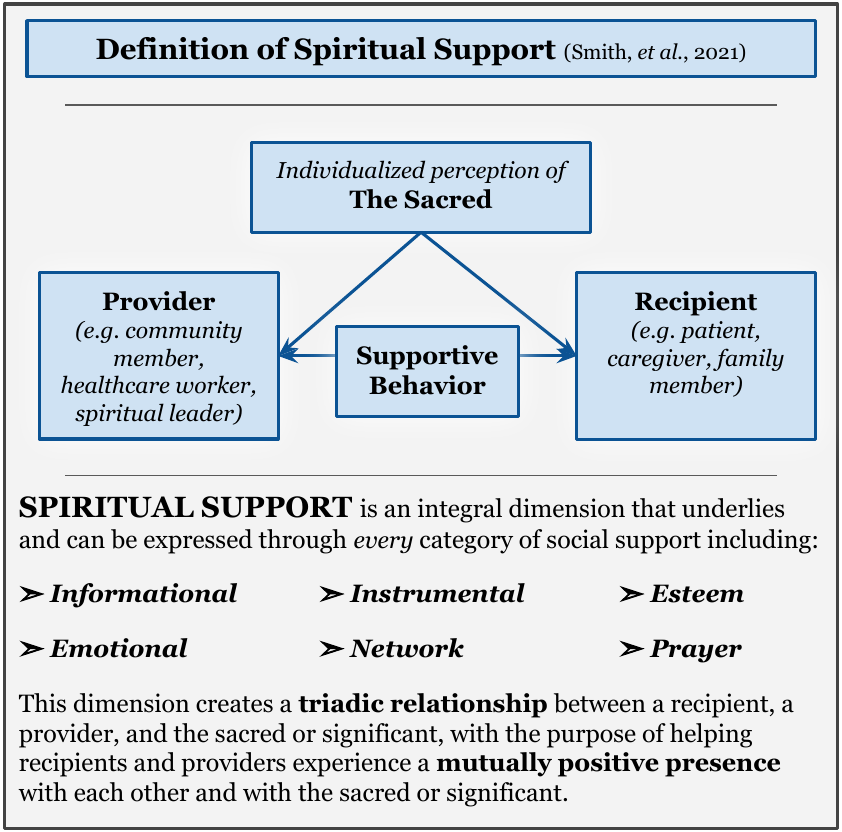}
    \caption{Definition of Spiritual Support from \cite{smith_what_2021}.}
    \label{fig:definition}
\end{figure}
}

\newcommand{\newdef}{
\begin{figure}
    \centering
    \includegraphics[width=\linewidth]{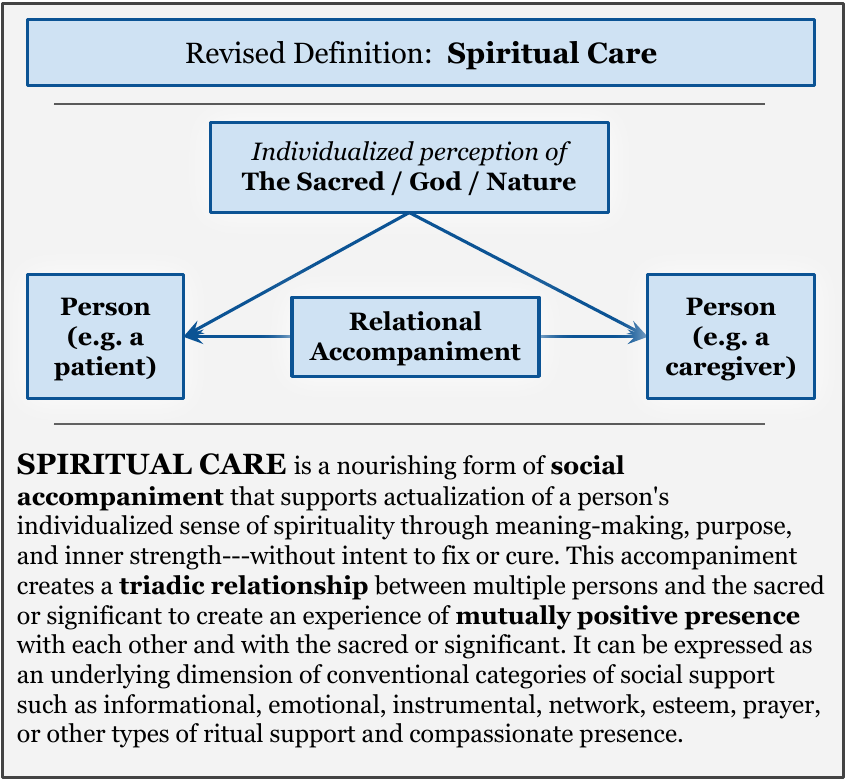}
    \caption{Revised Definition of Spiritual Care.}
    \label{fig:new-def}
\end{figure}
}

\newcommand{\participanttable}{
\newcolumntype{A}{>{\raggedright\arraybackslash}m{.5cm}}
\newcolumntype{Z}{>{\raggedright\arraybackslash}m{2.5cm}}
\newcolumntype{Y}{>{\raggedright\arraybackslash}m{2.8cm}}
\newcolumntype{D}{>{\raggedright\arraybackslash}m{0.7cm}}
\newcolumntype{B}{>{\raggedright\arraybackslash}m{1.8cm}}
\newcolumntype{C}{>{\raggedright\arraybackslash}m{1.1cm}}

\begin{table*}[t]
\footnotesize
    \centering
    \begin{tabular}{A  Z   Y   D   B   B   D}
    \hline
    \textbf{ID} & \textbf{Professional Title} & \textbf{Credentials} 
    & \textbf{Years} & \textbf{R/S Affiliation} & \textbf{Location} & \textbf{Reddit} \\
    \hline
    \hline
    SCP01 & Professor & Ordained Minister, Licensed Psychologist & \textgreater 10 & Christianity & Western USA & N \\
    \hline
    SCP02 & Chaplain & M.Div. & 1-5 & Islam & Southern USA & Y \\
    \hline
    SCP03 & Chaplain, Spiritual Leader & M.Div. & \textgreater 10 & Christianity & Western USA & N \\
    \hline
    SCP04 & Chaplain, Spiritual Care Counselor, Spiritual Director & M.Div. & 5-10 & Christianity & Western USA & Y \\
    \hline
    SCP05 & Spiritual Director, Spiritual Leader & M.S. in Pastoral \& Spiritual Care & \textgreater 10 & Christianity & Western USA & N \\
    \hline
    SCP06 & Director of Case Management & Licensed Social Worker 
    & 1-5 & Christianity & Western USA & Y \\
    \hline
    SCP07 & Chaplain & M.Div. & \textless 1 & Christianity & Western USA & N \\
    \hline
   SCP08 & Chaplain & Pending ACPE  
   & 1-5 & Humanist & Western USA & Y \\
    \hline
    SCP09 & Physician & M.D. & \textgreater 10 & Christianity & East Coast & N \\
    \hline
    SCP10 & Chaplain, Director, Spiritual Care \& Education & BCC, Advanced Practice BCC & 5-10 & Not Disclosed & East Coast & Y \\
    \hline
    SCP11 & Manager, Chaplaincy Department & M.Div., BCC, ACPE  
    & \textgreater 10 & Not Disclosed & East Coast & Y \\
    \hline
    SCP12 & Chaplain & BCC & \textgreater 10 & Christianity & East Coast & Y \\
    \hline
    SCP13 & Chaplain & BCC & \textgreater 10 & Islam & East Coast & N \\
    \hline
    SCP14 & Chaplain, Spiritual Director & M.S. in Buddhist Chaplaincy & \textgreater 10 & Buddhism & East Coast & N \\
    \hline
    SCP15 & Seminarian & Certified Death Doula & 1-5 & Buddhism & East Coast & N \\
    \hline
    SCP16 & Chaplain, Physician, Professor, Spiritual Care Specialist & M.Div., BCC & \textgreater 10 & Judaism & West Coast & Y \\
    \hline
    SCP17 & Chaplain, Professor & M.Div., BCC & \textgreater 10 & Judaism & West Coast & Y \\
    \hline
    SCP18 & Vice President of Spiritual Care & M.Div., ACPE 
    & \textgreater 10 & Christianity & West Coast & N \\
    \hline
    SCP19 & Chaplain, Mindfulness Educator & BCC & \textgreater 10 & Not Disclosed & West Coast & N \\
    \hline
    SCP20 & Chaplain & Doctorate in Ministry, BCC & \textgreater 10 & Christianity & Southern USA & N \\
    \hline
    SCP21 & Executive Director, Spiritual Care Counselor & M.Div., ACPE  
    & \textgreater 10 & Christianity & Southern USA & N \\
    \hline
    SCP22 & Chaplain, ACPE Educator & ACPE 
    & \textgreater 10 & Christianity & Western USA & N \\
    \hline
    \end{tabular}
    \caption{Participant Table. This table is replicated verbatim from~\cite{bezabih_meeting_2025}, except that participant identifiers have been altered from ``P'' in~\cite{bezabih_meeting_2025} to ``SCP'' (Spiritual Care Provider) to differentiate between the different stakeholder groups present in this paper. ``Reddit'' column indicates participants' self-reported prior familiarity (Yes/Y) or lack of familiarity (No/N) with the Reddit platform. M.Div. denotes a Master's in Divinity; BCC, Board Certified Chaplain; M.S., Master's of Science; ACPE, Association for Clinical Pastoral Education Certification.}
    \label{tab:participants}
\end{table*}
}

\newcommand{\dimensions}{
\begin{figure*}
    \centering
    \includegraphics[width=\textwidth]{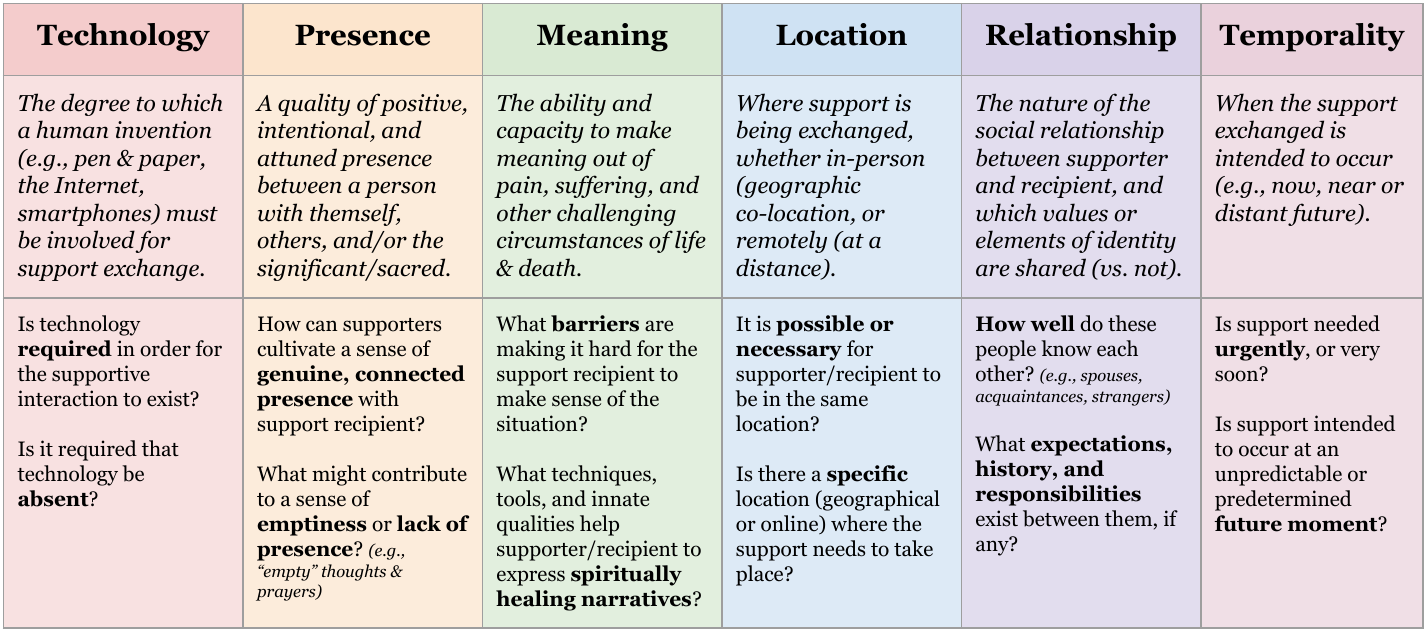}
    \caption{Design Dimensions for Technological Spiritual Care Interventions.}
    \label{fig:dimensions}
\end{figure*}
}

\maketitle

\section{Introduction}
Modern society is faced with epidemics of loneliness~\cite{health_oash_new_2023}, mental illness~\cite{kazdin_rebooting_2011,tucci_we_2017,american_academy_of_pediatrics_aap-aacap-cha_2021,substance_abuse_and_mental_health_services_administration_2023_2023}, and spiritual crises~\cite{wilt_struggle_2018,ferrell_urgency_2020}. The fields of HCI and medicine have focused predominantly on emotional and mental aspects of this issue, however spirituality and spiritual care are typically omitted or underdiscussed in the HCI research literature~\cite{buie_spirituality_2013,rifat_integrating_2022,smith_sacred_2022} and within patients' experiences of medical care~\cite{el_nawawi_palliative_2012,balboni_spirituality_2022}. Humans experience \textit{distinct} spiritual needs and problems during challenging or traumatic events and life-altering, -threatening, or terminal illness that are not well-addressed through mental and physical health care systems~\cite{puchalski_improving_2009}. Moreover, spiritual factors are correlated with human wellbeing~\cite{long_spirituality_2024}. Thus, neglecting spiritual issues already constitutes a painful unmet need, and it may also invisibly impede progress or compliance with other types of interventions for mental or physical health. 
Receiving professional spiritual care (esp. during palliative care) has been repeatedly demonstrated to improve important health and Quality of Life outcomes~\cite{balboni_provision_2010,balboni_spirituality_2022} and patient satisfaction with care~\cite{marin_relationship_2015} while also reducing overall costs of care~\cite{balboni_support_2011}. Yet there are not enough professional \textbf{spiritual care providers (SCPs)} available to meet growing needs. Similarly to mental healthcare~\cite{kazdin_rebooting_2011}, there is an urgent need to expand models of delivery for spiritual care interventions~\cite{ferrell_urgency_2020} and telechaplaincy~\cite{winiger_changing_2022,sprik_chaplains_2022}. Leveraging HCI research methods and interdisciplinary co-design practices will be invaluable for supporting the emerging expansion of spiritual care into digital spaces and technologies. Thus, this paper aims to provide guidance for future designers and researchers interested in tackling this challenging, nuanced, and impactful opportunity.

Apart from clinical settings and connotations of spiritual care, 
prior work in HCI has demonstrated that online platforms are capable of providing digitally-mediated forms of spiritual support. For example, a body of work conducted on the health blogging platform \href{https://www.caringbridge.org/}{CaringBridge} has demonstrated that users view the platform as a powerful site for exchanging prayer and other forms of spiritual and social support~\cite{smith_i_2020,smith_thoughts_2023}. Motivated by the vital clinical value of professional spiritual care~\cite{puchalski_improving_2009,balboni_spirituality_2022}, participatory design workshops were conducted with a variety of CaringBridge stakeholders in order to empirically derive a definition of \textbf{spiritual support} for HCI (Fig.~\ref{fig:definition})~\cite{smith_what_2021}. Although design implications presented in~\cite{smith_what_2021} were ultimately tailored to CaringBridge and other online health communities, the diverse participants and open-ended nature of the data collection in that study offers rich insights across broader technology domains.\footnote{For instance, the data also include tech-based opportunities for spiritual interventions related to human dignity, sound \& text, practical resources, VR/AR, AI, and physiology~\cite{smith_what_2021}.} In this paper, we acquired the original data from~\cite{smith_what_2021} to derive a more comprehensive design framework for technological spiritual care interventions beyond online health communities. 

Moreover, an essential element of interdisciplinary HCI work is to ensure that HCI researchers conscientiously and accurately integrate knowledge from technology stakeholders and external fields of study of relevance to the design questions at hand---in our case, spiritual care. Although a small number of chaplains were participants in~\cite{smith_what_2021}, the resulting definition of spiritual support has not yet been validated more broadly with members of the professional spiritual care community. Therefore, we also conducted new semi-structured interviews with professional SCPs ($n=22$) to discuss and refine the existing definition of spiritual support from~\cite{smith_what_2021} and its implications for HCI. This member-checking activity helps to ensure that HCI can align effectively with spiritual care in its future aims and pursuits. Synthesizing across these two data sources, this study addresses the following research questions:

\begin{description}
    \item[RQ1:] What are professional spiritual care providers' reflections on prior HCI work to define spiritual support?
    \item [RQ2:] According to spiritual care stakeholders, what guiding principles should inform the design of technological spiritual care interventions within and beyond clinical settings?
\end{description}

Our interviews with SCPs helped to refine the prior definition of spiritual support (Section~\ref{sec:RQ1}) and suggested a vernacular shift toward using the term spiritual \textit{care} to better align with the field. Combined with new data from SCP participants, our re-analysis of data from~\cite{smith_what_2021} also contributes principles to inform future design efforts (Sec.~\ref{sec:RQ2}). We found evidence of three prerequisite conditions that must be met in order for users to be able to experience meaningful spiritual care (Sec.~\ref{sec:prereqs}): openness to care; the perception of a safe space; and the ability to discern and articulate needs. We also identified six design dimensions that must be considered (Sec.~\ref{sec:designdimensions}): loving presence, meaning-making, appropriate degree of technology use, location, degree of relational closeness, and temporality. Through synthesizing these findings with prior related research, our discussion contributes a design framework for digital spiritual care interventions that we will refer to as \textbf{SPIRIT}~(Sec.~\ref{sec:SPIRIT}): \textbf{S}piritual \textbf{P}rerequisites and \textbf{I}nnovative dimensions fo\textbf{R} \textbf{I}ntegrating \textbf{T}echnology. The SPIRIT framework can help to establish \textit{what} digital systems for spiritual care should aim to accomplish and \textit{how} they should be designed through thoughtful consideration of prerequisites and innovative design dimensions.

\section{Related Literature}
\subsection{Differentiating Spirituality and Religion}
Religion and spirituality (R/S) are distinct but related terms: people can identify as religious, spiritual, both, or neither. For example, in United States,there is a rising proportion of "spiritual-but-not-religious" (SBNR) identifying individuals~\cite{rotolo_who_2023} and "religious nones"~\cite{smith_religious_2024} who are disaffiliated from religion. Approximately 22\% of Americans are SBNR, 21\% are neither spiritual nor religious, and  58\% are religious---some of whom consider themselves both religious and spiritual (48\%) and some of whom identify as religious-but-not-spiritual (10\%)~\cite{rotolo_who_2023}. Due to the central importance of these terms to our work, we will begin by clarifying the distinction. 

\paragraph{Spirituality} Numerous definitions of \textbf{spirituality} exist in the literature of different fields (e.g., religious studies, social sciences, etc.) In 2009, \citeauthor{puchalski_improving_2009} convened a group of health experts and faith leaders to derive the following definition: \textit{``Spirituality is the aspect of humanity that refers to the way individuals seek and express meaning and purpose, and the way they experience their connectedness to the moment, to self, to others, to nature, and to the significant or sacred.''}~\cite{puchalski_improving_2009} We focus on this definition because it has been widely adopted across healthcare fields and it has increased the scientific rigor of spirituality research and spiritual care in clinical practice. (This definition is also relevant because it was used in~\cite{smith_what_2021}, as we will describe in Sec.~\ref{sec:CB} below.)

\paragraph{Religion} Numerous definitions also exist for \textbf{religion}. For example, classical scholarly definitions in anthropology or religious studies from Durkheim, Spiro, and Geertz focus respectively on a \textit{``unified system of beliefs and practices''}~\cite{durkheim_elementary_2016}, an \textit{``institution consisting of culturally patterned interaction with culturally postulated superhuman beings,''}~\cite{spiro_religion_1966} or \textit{``a system of symbols which acts to establish powerful, pervasive, and long-lasting moods and motivations.''}~\cite{geertz_religion_1966} Religion is thus described as cultural, institutional, and fixed on specific sets of unifying beliefs or practices. 

\vspace{2mm}

Our work here focuses mainly on spirituality, which--as the definition above offers--is fully individualized (rather than a unified system or institution) and pertains to how a given person experiences connection to what matters most to them. For religious-identifying individuals, the word ``sacred'' (from the~\cite{puchalski_improving_2009} definition) may often be tied to their religion's conception of a "god," "deity," or other supernatural entity. As we will soon describe, spiritual care is \textit{not} intrinsically religious---however for some people, it does \textit{involve} religious beliefs, rituals, or practices.

\subsection{Professional Spiritual Care}\label{sec:spiritualcare}
To the best of our knowledge, there does not currently exist a precise, standardized definition of \textbf{spiritual care}; therefore, one of the contributions of this paper is to provide such a definition. Nonetheless, this section summarizes what the literature has to say about what spiritual care is and offers. 

According to seminal work by \citeauthor{puchalski_improving_2009}, \textit{``Spiritual care models offer a framework for health care professionals to connect with their patients; listen to their fears, dreams, and pain; collaborate with their patients as partners in their care; and provide, through the therapeutic relationship, an opportunity for healing.''}~\cite{puchalski_improving_2009} It is important to note that healing is distinct from physical cure. Rooted in spirituality, and using compassion and hopefulness, healing enables patients to find \textit{``solace, comfort, connection, meaning, and purpose in the midst of suffering, disarray, and pain,''} even with acceptance of the loss of social productivity and eventual death~\cite{puchalski_improving_2009}. The most common professional title associated with spiritual care is ``chaplain,'' however other titles such as spiritual counselors, spiritual directors, and faith leaders also provide spiritual care. In this paper, our participant samples include individuals with all of these titles who can consequently speak broadly on spiritual care provision. For concise reporting, we will use the acronym SCP ("spiritual care provider") to refer to our participants generically, and we will use the word "chaplain" only when we (or prior literature) refer to that specific professional title. 
It is essential to note that spiritual care tends to peoples' fundamental human needs for loving and being loved, belonging in community, and meaning and purpose~\cite{shields_spiritual_2015}, regardless of whether they affiliate with a religion. The newer acronym \textbf{SERT} ("spiritual, existential, religious, theological")~\cite{vieten_competencies_2016,rupert_clinical_2019} (as opposed to R/S) also captures "existential" concerns, since words like "spiritual" or "religious" can invoke strong, negative reactions from some people. Although delicate language issues can and will continue to challenge the field, spiritual care is intended to be accessible to \textit{any} person experiencing SERT-related struggles. We acknowledge that, similarly to any other form of healthcare, there may always be some individuals who choose not to seek or accept this form of care.

SCPs work across numerous contexts spanning clinics/hospitals, military, prison, educational and industry institutions, homelessness, and more~\cite{bezabih_meeting_2025}. In this work, we focus mainly on SCPs in clinical healthcare, where models of spiritual care typically assume a generalist-to-specialist form~\cite{hall_spiritual_2016}. This involves: (1) Clinicians asking for a patient’s spiritual history; (2) Assessment of current spiritual issues and needs; (3) Appropriate specialist referral(s) to trained and board-certified chaplains (in-patient) and/or spiritual care providers in the community (out-patient); (4) Development of a spiritual intervention or treatment plan; and (5) Continual re-evaluation of outcomes and spiritual needs during treatment~\cite{hall_spiritual_2016}. SCPs attend to a variety of crucial care tasks such as advance care planning and end-of-life directives and treatments~\cite{kwak_role_2021}, building relationships, ritual support (e.g., prayer, sacraments, etc.), connecting patients with communities of faith, attending to death/dying or grief/bereavement, addressing goals of care, and alleviating spiritual and existential distress through caring conversation~\cite{jeuland_chaplains_2017}. In alignment with clinical pathways used within US healthcare communication, the Advocate Health Care Taxonomy of Chaplaincy offers a rigorous delineation of ``intended effects,'' ``methods,'' and ``interventions'' that describe why, how, and what chaplaincy care precisely contributes to medical care~\cite{massey_what_2015}. However, unlike other healthcare professions, spiritual care is \textit{not} prescriptive; it does not view patients as needing to be ``fixed'' or ``cured.'' Rather, it provides patients with a point of human contact, care, and connection within a system that can feel otherwise chaotic, beyond one's own control or agency, and disregarding of an individual's humanity or dignity.

\subsection{Prior Work in Digital Spiritual Care}\label{sec:spiritualcaretech}

\subsubsection{Motivating the Expansion of Spiritual Care into Technology Domains}\label{sec:tech}
An imperative of patient-centered and whole person care is that spiritual care be equally available in both in-patient and out-patient settings, yet spiritual care is often unavailable to patients~\cite{handzo_chaplaincy_2022}. Reasons for this include insufficient clinical training in spiritual care, failure to include spiritual care providers on care teams, and lack of access in underserved populations~\cite{balboni_spirituality_2022}. More mental healthcare workers are now incorporating considerations of spirituality into their practice~\cite{schiffman_more_2022}, yet there are nowhere near enough practitioners to meet continually rising demand for mental healthcare as well~\cite{kazdin_rebooting_2011}. Mental healthcare is also distinct from spiritual care, in terms of its aims, training, credentialing, professional titles, and care contexts, as has been described in depth by prior work~\cite{bezabih_meeting_2025}. Prior to 2020, spiritual care had largely avoided  engaging in digitization, however the COVID-19 pandemic heightened the urgency of the need to consider new models of delivery for spiritual care~\cite{ferrell_urgency_2020}, as the next section will explore.

\subsubsection{Digitization of Spiritual Care}\label{sec:digitization}
The COVID-19 pandemic fueled trends in healthcare towards telehealth and virtual delivery of services, including in spiritual care and telechaplaincy~\cite{winiger_changing_2022,sprik_chaplains_2022}. Since SCPs could not sit with people in-person, \citeauthor{ferrell_urgency_2020} described a tragic lack of spiritual care for patients and their families and the absence of R/S rituals and funerals that help people make peace during and after a loss---both of which can have profound long-range consequences. Therefore, professional SCPs needed to rapidly pivot to remote spiritual care and to adopt digital tools and flexible workflows to maintain connection with patients and staff~\cite{szilagyi_covid-19_2022, snowden_what_2021, vandenhoeck_most_2021}. While this shift highlighted the potential for telechaplaincy to expand access, it also introduced distinct systemic challenges regarding the structure of digital interventions and the maintenance of professional standards in virtual environments~\cite{winiger_navigating_2024}. Furthermore, the long-term viability of these digital models depends heavily on user acceptability, which varies significantly based on patient demographics, prior technological experience, and the specific affordances of the hybrid or digital-only modalities employed~\cite{winiger_acceptability_2025}. 

The field of spiritual care is now continuing to grapple with how to effectively adopt technologies and design workflows for online care. For example, \href{https://telechaplaincy.io/events/}{Telechaplaincy.io} is a community of practice that hosts online events for networking and discussing emerging practices and strategies for telechaplaincy and digital spiritual care. However, a targeted framework derived from stakeholder input will greatly support the field to better design and embrace technology in useful and acceptable ways. We draw inspiration from recent work in the adjacent domain of mental health, where users must navigate and configure complex ecosystems of technology products, care providers, and social concerns~\cite{burgess_collaborative_2019,burgess_whats_2025,ongwere_challenges_2022}. For example, \citeauthor{slovak_hci_2024} propose using four levels of design briefs to guide intervention design: (1) technical capabilities (individual features or affordances of interaction design); (2) components (which incorporate one or more capabilities to deliver a particular design element); (3) intervention systems (which combine multiple components into a complete intervention); and (4) intervention implementations (which can be deployed with trained clinicians in specific institutional settings)~\cite{slovak_hci_2024}. We acknowledge that mental health and spiritual care are distinct professions with distinct aims, albeit with some overlapping care techniques (as is described in depth by~\cite{bezabih_meeting_2025}). However, we will borrow the language of these four levels in our discussion, since they usefully encapsulate similar design concerns shared between the two fields. 

\subsubsection{Designing for Religious, Spiritual, and Existential Concerns in HCI}\label{sec:CB}
Numerous design terminologies and approaches in HCI have emerged over time. For example, Norman introduced the original conception of ``User-Centered Design'' in 1986~\cite{norman_user_1986}, which has since evolved into ``Human-Centered Design''~\cite{interaction_design_foundation_what_2021}---an overarching framework that encompasses many methodological approaches such as participatory design~\cite{bodker_utopian_1987}, asset-based community design~\cite{mathie_clients_2003}, value-sensitive design~\cite{borning_next_2012}, value-sensitive algorithm design~\cite{zhu_value-sensitive_2018}, or human-centered machine learning~\cite{chancellor_who_2019,kaluarachchi_review_2021}. Cumulatively, these approaches seek to work closely with and empower communities of impacted technology stakeholders, identify peoples' needs, values, and assets, translate  findings into design implications, articulate major design risks/tradeoffs/inequities, and continuously evaluate the technology impacts following deployment. While SERT-related concerns and values \textit{can} be considered within existing frameworks, they have rarely been an explicit or central focus in decades of prior HCI research~\cite{buie_spirituality_2013,rifat_integrating_2022,smith_sacred_2022,ahmed_situating_2022}. In a seminal work that defines the term ``techno-spirituality," \citeauthor{buie_spirituality_2013} argue that SERT avoidance in HCI may be due to perceptions of risk that this topic could be viewed by peers as irrelevant, too difficult/sensitive, professionally risky, unscientific, or unfundable~\cite{buie_spirituality_2013}. However, avoiding SERT issues can marginalize (or even violate) needs and concerns that are of central and profound importance to \textit{most} of the world's population (and consequently, \textit{most} of the world's  technology stakeholders)~\cite{rifat_integrating_2022}. This both limits forms of care, meaning-making, or wellbeing that technologies might otherwise support, while biasing designs toward hegemonic (and supposedly ``secular'') values, rather than cultivating a pluralistic society where diverse worldviews can co-exist and thrive~\cite{rifat_many_2023}.

Encouragingly, recent years have seen a rise in HCI research publications related to SERT concerns and considerations for R/S in design~\cite{wolf_still_2024,kurosu_grounding_2025}. For example, some design frameworks are now proposing strategies for cultivating liminal~\cite{liedgren_liminal_2023}, soulful~\cite{halperin_miracle_2023}, or transcendent~\cite{buie_exploring_2018,buie_transcendhance_2016} user experiences. According to a recent review by~\citeauthor{kurosu_grounding_2025}, much of the literature focuses across four areas~\cite{kurosu_grounding_2025}: (1) designing technologies such as robots, tangible artifacts, and apps for specific religious traditions; (2) how technology is used in conjunction with R/S; (3) theoretical aspects of studying the connection between technology and R/S; and (4) providing background information on R/S practices to inform future design efforts.

\paragraph{Spiritual Care in HCI}Just as the literature on R/S is growing, there is also a small but growing body of work on spiritual care in HCI. Alongside patients' psychological, social, medical, and material needs, HCI research in palliative care has argued for the necessity to deeply consider patients' needs for spiritual care---for example, in the design of Virtual Reality (VR) experiences~\cite{ahmadpour_how_2023} or Embodied Conversational Agents (ECAs)~\cite{oleary_something_2024}. Chaplains have also been included as research participants to understand \textit{value elicitation} processes during palliative care (in which they are often central care team members)~\cite{doyle_navigating_2025}. To our knowledge, \cite{bezabih_meeting_2025} is the only prior work to focus exclusively on chaplain and SCP participants, who were consulted to understand whether and how they might extend their professional role to support users of online health communities. Results indicated that chaplains felt they \textit{should} participate online, either as community members whose contributions could model spiritually healthy norms for others or as moderators who could help to build and maintain safer and more trustworthy online care spaces. Through introduction of the ``Care Loop'' model, they also emphasized that online spiritual care should never \textit{replace} but rather \textit{supplement} in-person, professional care~\cite{bezabih_meeting_2025}. 

\ssdefinition

\paragraph{CaringBridge} CaringBridge (\href{https://www.caringbridge.org/}{www.caringbridge.org}) is a nonprofit online health communication platform that receives around 40M unique site visits per year. As demonstrated by a highly relevant body of prior HCI work~\cite{ma_write_2017,smith_i_2020,levonian_bridging_2020,smith_what_2021,ma_detecting_2021,ma_practical_2022,smith_thoughts_2023,smith_sacred_2022,levonian_peer_2025}, patients and caregivers turn to CaringBridge for sharing health information and coordinating social support during serious and life-threatening illness. For example, the majority of patients using CaringBridge have received a cancer diagnosis and approximately 36.9\% of cancer patient sites terminate with the death of the patient~\cite{ma_write_2017}. Thus, its context of use is highly adjacent to clinical (and sometimes palliative) care. In behavioral traces of users' journals to understand what types of support are exchanged on the platform, \citeauthor{smith_i_2020} found that appreciation for prayers received is the most common expression of appreciation, while survey results showed that users also rate prayer as the most important form of support exchanged on CaringBridge~\cite{smith_i_2020}. Given that prayer is often tied to R/S beliefs and practices, we sought to understand and define ``spiritual support'' in a follow-up study~\cite{smith_what_2021}. 

Using a participatory co-design methodology with  stakeholders including patients and caregivers, medical professionals (e.g., nurses, doctors) and chaplains, R/S community leaders, and CaringBridge employees, \citeauthor{smith_what_2021} derived the definition in Figure~\ref{fig:definition}~\cite{smith_what_2021}. Participants were shown the consensus definition of spirituality~\cite{puchalski_improving_2009} (see Sec.~\ref{sec:spiritualcare}) as a starting point, and were asked to define and discuss spiritual support. They viewed spiritual support as an \textit{underlying} dimension of other forms of support exchanged online--not the least of which is prayer. Given the prior evidence that CaringBridge has been a successful site for spiritual support, it provides a valuable initial model for inspiring future designs. The methods of this study seek to validate and refine this prior definitional work on spiritual support~\cite{smith_what_2021}.
\section{Methods}
Apart from the examples listed above, there is limited prior research in HCI conducted with professional spiritual care providers. Therefore, our study uses a Grounded Theory Method (GTM)~\cite{muller_curiosity_2014} approach to inductively derive a design framework grounded in empirical data for technological spiritual care interventions. We analyze data from two separate sources. We acquired original data from the design workshops conducted with CaringBridge stakeholders in 2019-20~\cite{smith_what_2021}. We then conducted new interviews with professional spiritual care providers (SCPs) to reflect on the prior definition. Both studies were deemed exempt from IRB review by ethics review boards at their respective institutions.

\subsection{Re-Use of Prior Co-Design Data}
In~\cite{smith_what_2021}, design workshops were conducted in late 2019 across four different stakeholder groups ($n=34$ participants): patients and caregivers (Participant IDs beginning with ``P''); spiritual or religious leaders of varying faith traditions (``L''); health and medical professionals such as nurses and doctors, and chaplains working in healthcare settings (``H''); and CaringBridge employees (``E''). Workshops lasted approximately 2 hours and participants worked in groups of 3 or 4 to first derive their group's own definition of spiritual support and next brainstorm many possible technology designs that could possibly facilitate spiritual support according to that definition. Finally, participants engaged in rapid prototyping activities such as sketching or using art supplies to build and discuss representations of their favorite ideas. (See~\cite{smith_what_2021} for extended protocol description.)

\subsection{Interviews with Spiritual Care Providers}
In order to assess the definition of spiritual support derived in~\cite{smith_what_2021} and to further discuss its implications for technology design, we conducted new semi-structured interviews with professional SCPs ($n=22$). In this section, we summarize our sample and data collection protocol. Participant IDs beginning with ``SCP'' indicate new SCP participants.

\subsubsection{Recruitment and Participant Sample}\label{sec:sample}
SCPs were recruited through a combination of outreach to our professional networks, snowball sampling, and cold emails. Eligibility criteria included adults 18+ years of age with professional training or credentialing in spiritual care and/or chaplaincy, which yielded a diverse sample of providers (see Table~\ref{tab:participants}). All participants have professional credentials or certifications such as Masters of Divinity (M.Div), Board Certified Chaplain (BCC), Accredited Clinical Pastoral Education (ACPE), Licensed Social Worker (LSW), etc. Most participants provided \textit{multiple} relevant yet distinct professional titles, showing that they are experienced across different areas of spiritual care: 14 listed "chaplain" and reported practicing in clinical or community healthcare settings (only 6 listed \textit{exclusively} the title of chaplain); 5 work in spiritual care management (e.g. Director of a Spiritual Care Unit), 4 are spiritual care educators (e.g., professors or ACPE instructors); 3 work as spiritual counselors or directors); and 3 work as R/S community leaders. We intentionally sought participants of diverse R/S backgrounds, geographies within the USA, genders, and age.\footnote{14 participants self-identified with Christianity (55\%), 2 each (9\%) with Islam, Judaism, and Buddhism, and 1 with Humanism (atheist); 3 participants did not disclose R/S background (14\%). All participants were in the USA, including 4 on the West Coast (18\%), 8 Western USA (36\%), 3 Southern USA (14\%), and 7 East Coast (32\%). 11 participants self-identified as female (50\%), 10 as male (44\%), and 1 as non-binary (5\%). Most participants are highly professionally experienced and advanced in age. 15 (68\%) report having >10 years of professional experience, 2 have 5-10 years, 4 have 1-5 years, and 1 has <1 year of experience. For age, 2 participants are 25-34 years old (9\%), 4 are 35-44 (18\%), 5 are 45-54 (23\%), 5 are 55-64 (23\%), and 6 are >65 (27\%).} Per GTM best practices, recruitment continued until data saturation was achieved---i.e. even though new participants discussed new specific examples, they were no longer raising new concepts~\cite{muller_curiosity_2014}. Participants received a \$50 eGift Card as compensation. (See~\cite{bezabih_meeting_2025} for full protocol, including interviews and user testing sessions.)

\participanttable

\subsubsection{Interview Protocol}
Interviews were the initial portion of a larger research protocol that subsequently involved user-testing sessions for a total session lasting approximately 2 hours/participant. Interviews were about 30-40 minutes of the full session. This paper reports \textit{only} on the interview data of relevance to the research questions of this paper; the specific methods and results of the remaining user testing portion of these sessions are reported in~\cite{bezabih_meeting_2025}. 16 sessions were conducted in-person (73\%), with the remaining 6 over Zoom (27\%). In advance of each session, participants completed a brief survey to provide demographic information and informed consent. During semi-structured interviews, the questions first focused on participants' regular professional activities and how they do or do not integrate technology in their spiritual care practice or organization. We then described prior work on CaringBridge and showed them a print-out of the definition of spiritual support from~\cite{smith_what_2021} exactly as it appears in Figure~\ref{fig:definition}.\footnote{In order to avoid confusing participants about what we wanted feedback on, we did not \textit{also} show them \citeauthor{puchalski_improving_2009}'s definition of ``Spirituality''~\cite{puchalski_improving_2009}. However, during their interviews, many participants specifically brought up \cite{puchalski_improving_2009} as a point of comparison or aspect of their background training.} We requested that they share any reflections on this definition, as well as sharing prior experiences (if any) related to technology use in their practice, as it relates to the presented definition. Interviews were recorded and automatically transcribed using Microsoft Word. Transcripts were manually double-checked and corrected against the audio recording. De-identified transcripts have been deposited to the Qualitative Data Repository (QDR) at \href{https://doi.org/10.5064/F6R7J9HL}{https://doi.org/10.5064/F6R7J9HL} for open data re-use or re-analysis in accordance with our ethics board-reviewed protocol.

\subsection{Analysis}
This work is part of an overarching Grounded Theory Method (GTM) approach~\cite{muller_curiosity_2014}, which can take years of thoughtful, iterative analysis to perform well. GTM is methodologically appropriate since both the fields of HCI and spiritual care do not currently have well-established frameworks for online spiritual care, and GTM offers a way to inductively derive foundational theory and frameworks through rigorous data collection, analysis, and data-driven modeling. GTM involves ``constant comparison'' of incoming data and results against data collected earlier in the process in order to refine emerging theory or models~\cite{muller_curiosity_2014}. \citeauthor{smith_what_2021} began this process in their original paper on ``Spiritual Support'' from 2021~\cite{smith_what_2021}. Our work in this paper continues it both through comparison of older data with new data and member checking to ensure validity and desirability to impacted stakeholders.

We began analysis by inductively open coding all data, writing memos, and conducting affinity mapping of the open codes using Miro boards to identify axial themes \textit{separately} for both data sets. Next, we narrowed the scope of our analysis by identifying clusters across both of the affinity maps that were of relevance to our research questions. More specifically, we collated and re-clustered axial themes from the prior workshops with CaringBridge stakeholders related to prerequisites or design dimensions for spiritually supportive technology designs---including but not limited to online community design. We then selected clusters from our affinity map of new interview data that described chaplains' perspectives on the spiritual support definition. Participants \textit{did} offer critiques and refinements of the spiritual support definition from~\cite{smith_what_2021}. Therefore, to serve our goal of member-checking with the spiritual care community, we report these in Sec.~\ref{sec:RQ1}. Finally, we reviewed axial codes related to other aspects of chaplains' regular practice and use (or non-use) of technology. We triangulated these data with the clusters identified in the CaringBridge data on prerequisites and design dimensions by adding relevant open codes from the chaplains' new data to the existing clusters. Consequently, the data reported in Sec.~\ref{sec:RQ2} are synthesized across both data sets. 

Using the presented results, our team inductively derived a new framework for designing technology to support spiritual care, which is presented in the discussion. To finalize our member checking objective, we sent the paper manuscript to all of our new SCP interview participants for any final refinements or revisions prior to publication. This step ensures that our work can be respectful of and useful to the professional spiritual care community.
\section{RQ1 Findings: Definitional Refinement}\label{sec:RQ1}
Participants appreciated that the provided definition~\cite{smith_what_2021} was a good first attempt---particularly the triadic concept and the aspect of mutuality in spiritual support. However, they also offered nuanced critiques, which we focus on in this initial results section. First, participants emphasized that a broader understanding of \textit{who we are} as humans is a prerequisite for defining spiritual care (Sec.~\ref{sec:broader}). Second, they suggested alternative ways of understanding (or replacing) other key terms in the provided definition, such as ``support'' and ``provider'' (Sec.~\ref{sec:keywords}). Finally, we offer a revised definition synthesized from participants’ reflections (Sec.~\ref{sec:revised}). 

\subsection{Broader Perspectives on Humanity and Spirituality Determine the Subject of Care} \label{sec:broader}
Before engaging with the presented definition of ``spiritual support'' from~\cite{smith_what_2021}, participants tended to first reflect on broader perspectives about the nature of humanity. For example, given the current global polycrisis, SCP16 emphasized the importance of \textit{``understanding \textbf{who we are} as human beings. What does spiritual care for a changing world mean?''} Participants suggested that a better understanding of human spirituality determines what is the \textbf{subject} of care and \textbf{how} to provide that care. They described two aspects of human spirituality that must constitute major subjects of spiritual care---i.e., spirit (inner-self) and relationships---and they wished to see both aspects reflected in a revised definition.

\subsubsection{Caring for the Inner-Self} \label{sec:innerself}
According to SCP13, \textit{``everyone has a spirit,''} and spirituality is personal and within a person; it cannot be provided from the outside but must be self-discovered from within. Yet external disturbances--such as bodily pain--can affect people's spirit, feelings, and actions. Thus, the first role of spiritual care is to help people search, care for, and realize the spirit and inner strength that is already within. 
SCP14 discussed the incomprehensible nature of ``spirit'' or ``inner self,'' and explained it in an \textit{apophatic}\footnote{\textit{``Apophatic''} is a common term in theology. It means that an unknowable essence of God is better explained through negations (e.g., God is not a body) than using positive statements (e.g., God is [something]).} way: \textit{``It's very unscripted, it's unknown. It's not linear. It's not even a triad of things, it's really to me, and this is a Buddhist [perspective] in a way, but it's a process of entering into presence.''} Due to the incomprehensibility of spirituality, SCP14 
said, \textit{``this [current definition] is very clean and very thoughtful. But I wish it was \textit{messier}.''}

SCP7 described caring for the inner-self/spirit as helping people to \textit{``improve how they experience themselves as sacred in the world.''} When it comes to a time of spiritual distress (e.g., caused by a physical disease), SCP17 said, \textit{``we ensure that the inner life is included in the care.''}  
Participants noted that everyone has the spiritual potential to cope with challenges, but rediscovering it needs support. For example, SCP13 uses the metaphor of a burning home to help people positively accept and cope with challenging situations:

\begin{quote}     
      Your house catches fire. What's the first thing you're going to do? You're going to try to put it out with water. And if the water doesn't put it out, you call the fire department. And if the fire department is not able to put it out, what do you do? You leave the house and go live in a new house. The same thing with our bodies and our spiritual life. That spirit, that life inside, is in the house. And when this house breaks down, ... you got some big great doctors, hospital systems, nurses, but they can't keep your house from burning down. It's not \textit{whether} you leave, it's \textit{how} you leave. ... You can have a state of mind that can be really competent, comforting, and make you feel you're in heaven.
\end{quote}

This rediscovery of one’s inner spiritual strength is not something that can be given externally. Thus, participants suggested that spiritual care should be understood as a process of facilitating self-rediscovery---i.e., inner work that grants greater agency to the recipient---rather than as something "bestowed" by external helpers.

\subsubsection{Caring for Relationships}
Another dimension of humanity considered by participants especially relevant in spiritual care was that people are \textit{relational}. As SCP16 put it, \textit{``We are social beings, so there's that social which is relational and we belong to each other. We're part of constellations of relationships and ecologies, and then also institutions and cities and discourse.''} Therefore, caring for relationships constitutes the second core subject of spiritual care. 

Regardless of peoples' R/S backgrounds, participants talked about two types of relationships: (1) Relationship with the Divine, Sacred, or Nature, (2) Relationship with Others. As SCP16 described, \textit{``who are you in relationship to others, who are you in relationship to all that has life, to the cosmos, to God, to whatever, is of the Ultimate to which we all belong.''} Participants acknowledged that what is considered divine or sacred can be diverse depending on peoples' R/S backgrounds: \textit{``whatever the person or community holds sacred.''} (SCP11) They also reflected that almost everyone, even an atheist, has a sacred ``thing'': \textit{``Seculars might not believe in God or they may not be religious, but very few people would say there's nothing sacred to them.''} (SCP11) The role of spiritual care then is to facilitate this \textit{``connection with the divine, whether that's God, Buddha, whatever their higher power is.''} (SCP2) Spiritual care also considers the communal aspect by facilitating \textit{``connectedness and belonging''} (SCP16) to others and community. 

\subsection{Revising Key Words in the Definition}\label{sec:keywords}

\subsubsection{The Word ``Support'' is Transactional}
Rather than the word ``support'' in the provided  definition from~\cite{smith_what_2021}, participants preferred the word ``care.''  They felt that ``support'' connotes a sense of transaction rather than a spirit of service. For them, care is \textit{``really about connection''} (SCP14) and \textit{``constitutive of who we [humans] are''} (SCP16) as everyone cares. SCP16 also highlighted that support suggests a curative perception or a fixing of others' condition through some kind of authority, but spiritual care aims to instead provide compassionate presence: \textit{``I'm bringing a quality of presence to the moment, no matter what's going on for you. ... That's different than helping. For me, this is what it means to be sacred.''}

\subsubsection{A Spiritual Care Provider is a Compassionate Friend, Not an Authoritative Fixer}
Participants underscored that, in spiritual care, the provider-receiver relationship should be understood as mutual/reciprocal, happening among ``equals.'' They recommended a perspective where the provider (whether a professional or layperson) is seen as a compassionate friend, not an authoritative figure. This relationship (whether professional-to-patient or peer-to-peer) should always be perceived as bi-directional, even when one party is seemingly in more crisis. SCP12 described this as follows:
\begin{quote}
    Each of us in our humanity brings something healing to others we interact with. So even a person in bed might be providing a healing presence to the person who is their caregiver. If we think of spiritual support exclusively as care provider and care recipient, I would want to push back on something that locks somebody into one or the other of those.
\end{quote}

Participants particularly stressed the need to be humble and avoid a \textit{``savior mentality.''} (SCP13) Instead, SCP3 highlighted a need for openness and voluntary vulnerability: \textit{``I'm not here to save the day. I'm here to be vulnerable and open with another person, and allow them a space to be vulnerable. That's safe, so that we might both experience something that's outside of ourselves while also having a very human interaction.''}

\subsection{Revising the Definition}\label{sec:revised}
Based on sections~Sec.~\ref{sec:broader} and~Sec.~\ref{sec:keywords}, it is evident that the original definition from~\cite{smith_what_2021} is missing important concepts. Rather than ``Spiritual Support,'' we offer the following revised definition of ``Spiritual Care'' that merges components from the new data with the original definition:

\begin{description}
\item[\textbf{Spiritual Care}] is a nourishing form of social accompaniment that supports actualization of a person's individualized sense of spirituality through meaning-making, purpose, and inner strength---without intent to fix or cure. This accompaniment creates a triadic relationship between multiple persons and the sacred or significant to create an experience of mutually positive presence with each other and with the sacred or significant. It can be expressed as an underlying dimension of conventional categories of social support such as informational, emotional, instrumental, network, esteem~\cite{cutrona_controllability_1992}, prayer~\cite{smith_i_2020}, or other types of ritual support and compassionate presence.
\end{description}

The messy reality is that spiritual care may still defy idealized academic description, but this revised definition is better aligned with SCPs' views. It  replaces the old definition's emphasis on ``provider'' versus ``receiver'' with an emphasis on ``persons,'' and it touches on both a person's \textit{inner self} and on external \textit{relationships}. Although spiritual care is a mutual experience, in the remaining findings, we will occasionally use the term ``receiver'' to refer to persons in distress or in need of care (e.g., patients) and the term ``provider'' to refer to supportive persons (e.g., chaplains or other healthcare providers, family, friends, etc.) to help with concision and clarity of our communication, while also acknowledging the limitations of imperfect language.
\section{RQ2 Findings: Prerequisites and Design Dimensions}\label{sec:RQ2}

\subsection{Prerequisites}\label{sec:prereqs}
Our data suggest that there are at least three prerequisite conditions that must be met before individuals can have an experience of spiritual care. Each condition may also be hindered from being met by certain barriers or challenges. 

\subsubsection{Openness to Care.} 
To participate in an exchange of spiritual care, both persons need to be flexible and open to the possibility of sharing meaningful moments of care. As H6 said, \textit{``what precedes the ability to seek and express meaning and purpose [is] being open to it.''} Participants also described how issues like the desire to control uncontrollable outcomes, denial of illness and death, and reluctance to re-adjust beliefs in light of a new reality can blind people to the possibility that they need or would benefit from care. For example, H7 described how people who belong to a faith community \textit{``sometimes do worse because there's not just a physical response to the injury, there's a maladaptive spiritual response---`Why is God doing this to me?'''
} 

Instead, flexibility \textit{``to readjust their spirituality'' (H7)} and \textit{``grace to accept things as they are'' (H1)} can create a genuine openness to support. For instance, P11 (who has battled a rare and incurable cancer for 20 years) said, \textit{``When we are able to release that tight death grip on thinking we can control things, that's when we are in a state of 
letting what's greater than me enter and have more power.''}  Equally important to the receiver's ability to accept support is the provider's ability to \textit{``attenuate''} to the receiver and \textit{``dial in to whatever frequency that person is broadcasting on.'' (H4)} Consequently, barriers to being an effective carer include trying to offer advice (rather than listening) or to impose one's own beliefs on another person.

\subsubsection{Safe Space.}
Participants often emphasized the necessity of cultivating a safe space prior to spiritual care---for \textbf{all} persons involved. For example, L5 described a time when her partner was hospitalized, \textit{``The pastor was in the office and it was beautiful because I knew he was just holding a safe space for me.''} Since providing continual care can be physically and spiritually draining for family and professional caregivers, participants like P11 called for care providers to \textit{also} have safe spaces for preparation and recuperation, \textit{``otherwise they're going to run dry.''} Unfortunately, H5 explained that many people don't receive training to care for their own needs, and H7 said, \textit{``No matter how good you are, you can only fake it for a while.''} Participants believed social media or other tech could potentially offer safe spaces for both persons in spiritual care exchanges, but they also expressed concerns about ``negative energy'' online. For example, P3 said, \textit{``CaringBridge is really specific to what people are going through, and I think that would be easier for people to feel that spiritual connection more than let's say Facebook. CaringBridge seems like a safe space.''} 

\subsubsection{Discerning and Articulating Needs.} During health crises, it is often difficult for individuals to look within themselves and discern what they need. For example, L2 said people are often \textit{``clueless''} while P5 said, \textit{``sometimes they don't even know what they need until you bring it up.''} If a person cannot express their needs, others may be in a difficult position to respond. Thus, P10 described how one critical prerequisite is \textit{``helping people to articulate what they need or what they want, and being okay with that vulnerability.''} However, barriers to discerning and articulating needs can include individual personality traits such as pride, independence, and stubbornness. For example, P6 said that articulating needs is incredibly hard for him because \textit{``I am stubborn and I would probably be like, `I am fine.' Then it would take others to force it out of me.''} Other barriers are more situational, including distraction due to extreme physical pain, such that \textit{``the crisis is currently stamping out the fire, so that distracts them from their ability to articulate the need for spiritual support.''} (H7) H4 also described how most people also experience serious psychological and existential distress, \textit{``Often what [patients] have inside of them is fear and anger and anxiety and hopelessness. And those are really hard things to share.''} 

\subsection{Design Dimensions}\label{sec:designdimensions}
Participants' discussions suggest six design dimensions that characterize the exchange of spiritual care, most of which can be interpreted as a continuous spectrum. We will focus on the two ends of each spectrum, however experiences can fall anywhere between them. The single most frequently and emphatically discussed concept was that of ``presence.'' Therefore we describe this aspect first and in the most depth.

\subsubsection{Loving Presence.} ``Being present'' refers to a person's own mindfulness and self-awareness in each moment, as well as their ability to intentionally set aside their own ego to be fully \textit{with} another person. A great deal of work in HCI work has examined ``presence'' in terms of using technology to overcome barriers associated with users being located in different places (e.g. video chat helps users feel more present with each other than email). We differentiate our usage of the word here to instead refer to an \textit{energetic quality} associated with a person's affect, behavior, and state of mind during an exchange of spiritual care. This energetic quality is intuitively discernible whether or not persons are physically separated. Although SCP5 worries about how, online, \textit{``there's so much lost''}, SCP16 asserts that \textit{``quality presence can happen even when virtual ... over different media or kinds of connectivity. In person, there's something energetically that you actually feel body-to-body. But also, like when you speak, it touches me, whether it's in person or on Zoom.''} SCP13 describes this ``calming'' while L2 calls it a \textit{``non-anxious presence.''} SCPs often focused on a non-judgmental affect that avoids any sense of trying to ``fix'' people. As SCP3 put it, \textit{``Fixing is judgment. Like, we fix broken things. We're not called to fix, we're not called to even to help. We're called to serve.''} For H4, cultivating a calm, connected presence means listening deeply, which requires only \textit{``two good ears, a wondering mind, and a caring heart.''}

When people are physically co-located, this type of intentional and beneficial presence is communicated through body language, posture, eye contact, and when needed, the willingness to be silent and simply \textit{be} together. \textit{``You just are with them, and you don't have to say much.'' (P11)} When people are physically separated, participants discussed how this quality of intentional presence is primarily communicated verbally (or via other media and online platforms) through genuinely thoughtful words and gestures. Interestingly, several participants also described supernatural experiences, when they simply felt a sense of presence, even without directly hearing from the other person through any form of technology. For example, L5 shared a story about feeling her friend's presence in a way that was \textit{``massively influential, outside of time and space''} the moment after the friend had been in a bad car crash.

On the other hand, L2 used the term \textit{``non-presence anxious''} to refer to a person whose affect is frenzied, distracted, or fixated on their own experience, feelings, and opinions/advice rather than the other person. L5 described this as \textit{``empty.''} In many cases, people are not being intentionally hurtful, but they lack the self-awareness or internal preparedness to be a positive presence. As a result, they do or say things that inappropriately impose their own beliefs, avoid major issues or conversations (e.g. acknowledging death), or focus on their own problems and pain, all of which can be damaging to others. Referring to in-person scenarios, P2 said, \textit{``It can be scary to see someone in the hospital bed being sick. Sometimes, the people who bail, or who are not there unless somebody forces them to stay there, sometimes they can drag the spirit down.''} Yet this can also be tangible online. Based on her experience of people's personalities, L2 talked about how, when reading Facebook comments online, she felt that some users genuinely \textit{mean it} when they say \textit{``I'll pray for you,''} while for others, they are \textit{``empty prayers,''}  typed from a sense of duty or convenience or normative compliance, rather thinking of something more caring and thoughtful to write. Likewise, SCPs felt that Reddit comments were often empty---nothing more than quick, shallow advice rather than actual care. For example, SCP2 offered a distinction between replying in a \textit{``dutiful and stewarding manner versus just like, OK, let me just give my two cents.''}

\dimensions 

\subsubsection{Meaning-making.} Spiritual care can help people make meaning of their situation, either by finding ways to outwardly express their own meaning, or by sensing meaningful moments being offered by others.
For SCP19, one aspect of a chaplain's role is to \textit{``support a patient in making decisions that are based on how they seek and express meaning and purpose, and all the ways that they do that, the sacred, and so on.''} For example, H2 described working with an anorexic woman near death who was refusing a feeding tube because she felt like a failure for being unable to follow her food plan. Through deep listening, H2 helped her realize that the feeding tube could bridge her back to the plan. \textit{``It helped her make sense of it, and then she was good with the feeding tube. Having the conversation helped her explore her meaning.''} While H2's story revolved around conversation, expression can also occur through other media. For example, P3 uses writing to express meaning. \textit{``I have been coming to terms with my cancer diagnosis and being able to connect with people through CaringBridge, me being able to talk through it through my writing and then having their comments come back.''} P11, a longtime painter, also highlighted the power of artwork as an expressive tool, \textit{``Somebody said, well why don't you paint yourself completely healed? That was a gateway for me to learn some amazing things.''}

Participants also discussed a variety of ways to help people \textit{sense} meaningful moments of connection, e.g. through saying prayers, playing music, singing, pleasant smells/tastes, or touch. For example, L3 was once a chaplain in the burn unit, \textit{``I remember the doctor saying that the one thing you have to do is find a way to touch these people. Because when \textbf{we} touch them it hurts. \textbf{You} need to just touch them in a way that makes them feel human without all this pain.''} Many participants' stories involved holding hands while speaking quietly or sitting in silence with another person.

\subsubsection{Appropriate Degree of Technology Use.}\label{sec:appropriatedegree} Participants repeatedly described technology as a paradoxical solution. As L5 said, \textit{``Social media is a black hole that can really cause a lot of people to be sick. At the same time, it can be very powerful.''} L2 similarly reflected, \textit{``With technology, the gift is you can go broader, you can do quicker, but also that's the problem with it, you can go too quick without thinking.''} Using technology for spiritual care is not only intrinsically a delicate balancing act, but there is also a spectrum for how much technology \textit{should} be used in specific contexts.

There are cases when participants found it completely inappropriate for technology to be providing care. For example, H4 said, \textit{``God help us if it's just robots taking care of people at end of life.''} As people are dying, L1 suggested that nothing can replace the in-person, physical presence of care providers. During her father-in-law's hospice care, she said, \textit{``Everyday I would go and sit and pray with him, and that was just the highlight of his day. No technology involved.''} There are other pragmatic cases when technology is neither desirable nor feasible. For instance, SCP11 shared that ``for inpatient hospitalized people, low tech does it best.'' Likewise, for people in the Intensive Care Unit, SCP17 shared, \textit{``there's no way they're going to look [at technology], not only just read something or look at something small or even look at something on a computer. They wouldn't be able to do that.''}

However, there are also instances when technology facilitates support that could not be possible without it. Recalling an inability to see patients in-person during the COVID-19 pandemic, SCP14 described using phone calls instead. \textit{``It's not the same as holding their hand, but it's holding their hand the best we can.''} Other participants described how technology provides different formats of care that feel profoundly meaningful in new ways. For example, L5 shared, \textit{``Being able to put a message out to your entire community, everyone's connecting, you have that many people synced in their intention, that's just that much more energy that's going in that direction of support and healing.''} And on the receiving end, participants described feeling that energy in powerful ways---such as H1, who shared that receiving CaringBridge comments \textit{``is like riding the wave of love. Both of us could feel this support that was in the writing.''} 

Ultimately, careful discernment is required to determine both the \textit{appropriate degree} of technology involvement, and how that technology is designed. As SCP14 put it, \textit{``It's not \textbf{whether} we use these platforms or not, because it's going to happen. It's more like, \textbf{what} are the platforms embodying.''} And as SCP15 noted, technologists can innovate new techniques for effective engagement, \textit{``You can sort of design spaces that are, yeah, a little bit more creative and targeted and focused. I've seen it to be an effective medium.''}

\subsubsection{Location.} Technologies for spiritual care need to consider geographical location, since care providers may be located anywhere from the same room to a different continent. \textit{``It's one thing if you're waiting in the lounge after surgery, but if somebody's in the hospital for weeks on end, their community can't be there because they've got their own lives.'' (H7)} For example, SCP16 shared a story about a daughter sitting bedside with her dying father. \textit{``There was somebody zooming in from FaceTime who was in Jerusalem at the time making prayers for him at the Western Wall.''} The real-time connection with this remote family member transcended distance to bring a faraway sacred space and experience to this father's final moments as a profound act of care. P14 raised an interesting point that providing care across distance can actually \textit{re-emphasize} the spiritual nature of connection. \textit{``I found that not having the physicality actually enhanced the truth of the interconnection. It wasn't dependent on form. It wasn't dependent on structure. It was dependent on \textbf{heart.}''} Finally, some SCPs were eager to improve access to care for target populations such as homeless, immobile, working, or other disadvantaged persons. For example, SCP7 wants to open their own online care platform in the future \textit{``that could be used to help people have more access to these type of services.''}

\subsubsection{Degree of Relational Closeness.} Participants discussed  how well care can be provided based on how well people know each other or what they hold in common. Care can come from anyone from unknown strangers to dearest family members. It doesn't always make sense to get support from strangers, for example \textit{``If some stranger texts me to say `I'm praying for you,' I'm like, you don't know me.'' (L2)} However, H4 (and many others) expressed that, \textit{``Oftentimes, [support] comes as a surprise and it comes from places that we didn't necessarily expect.''} For example, P3 noticed that strangers began visiting their CaringBridge site, \textit{``They don't comment, but I see that they just follow along and that is really neat to me that I am connecting with people, because they come back again and again.''} SCP11 shared that, even if people are strangers, they may be able to connect deeply if they share certain identity aspects or some common human experience, \textit{``...not just like a topic, but like a real feeling of deep heart, soul level [connection]. That's where people really get support that is interesting to see. Like, go to the baseball game, this kind of music, or whatever.''} At the same time, participants also shared that there are times when it feels absolutely vital for specific, known individuals to be available, such as the patient's favorite priest, nurse, family member, or friend---no one else could \textit{possibly} do.

\subsubsection{Temporality.} 
Participants emphasized that temporality is an essential consideration. At one end of the spectrum, support is sometimes required \textbf{immediately}. For example, P9 was feeling suicidal and called a good friend who said to her, \textit{``Can you just stay alive tonight? Just tonight?''} As SCP4 emphasized, \textit{``Sometimes it's just one person just reaching out''} that can make the difference between life and death. On the polar opposite end of the spectrum, support can be intended for legacy work or distant future generations. P3 shared, \textit{``I feel like, to have my thoughts and my words continue on beyond me is a part of keeping my spirit alive.''} Participants related this notion to the concept of sacred texts as technologies that continue providing many people with spiritual care. P4 shared an example about reading from the bible, \textit{``it is somebody's words from hundreds, thousands of years ago, who was speaking to you in this time when you needed support.''} As P9 said, \textit{``Words can carry these ideas through space and time.''} 
\section{Discussion}\label{sec:discussion}
Our study refines the definition of spiritual support from~\cite{smith_what_2021} with input from professional spiritual care providers (SCPs), leading to an updated definition (Figure~\ref{fig:new-def}). This research also uncovers three prerequisite conditions (Sec.~\ref{sec:prereqs}) and six design dimensions that are crucial for the effective delivery of spiritual care through technological interventions (Sec.~\ref{sec:designdimensions}). In this discussion, we synthesize the results to provide a design framework for implementing sociotechnical processes or interventions aimed at enhancing spiritual health. First, we describe the revised definition (Sec.~\ref{sec:definitiond}). Next, we present the SPIRIT framework (Sec.~\ref{sec:SPIRIT}) and provide a hypothetical example of how the framework could be applied (Sec.~\ref{sec:example}). 
Finally, we acknowledge limitations and the need for future work (Sec.~\ref{sec:futurework}).

\subsection{Revised Definition of Spiritual Care}\label{sec:definitiond}
\textbf{How can HCI researchers and technology innovators build new tools and systems that support meaningful spiritual care processes or interventions?} First, it is essential that researchers and designers understand what spiritual care is (and is not) in order to clarify misconceptions and develop a shared understanding with spiritual care stakeholders. To support this goal, our study refines a prior definition based on insights from professional SCP participants. Prior work used the term \textit{spiritual support}~\cite{smith_what_2021}, which implied a transactional model where care providers, positioned as experts, offered solutions to those in need. This framing overlooked the fundamentally relational nature of spiritual care. Drawing from SCPs' experiences, the revised definition of spiritual care presented in Figure~\ref{fig:new-def} centers on \textit{accompaniment} and \textit{compassionate presence}. Spiritual care is not about \textit{fixing} someone or providing prescriptive answers; rather, it involves creating a safe and empathetic space where individuals feel heard, valued, and accompanied in their meaning-making process. Moreover, the earlier definition missed two critical dimensions of spiritual well-being. First, it did not adequately account for the role of a person’s \textit{inner self}, including personal reflection, resilience, and hope. Second, it overlooked the influence of \textit{external relationships}—connections with divine, sacred, nature/ecology, family, community, and broader social structures that shape spiritual identity and care needs. By integrating these perspectives, our revised definition offers a more holistic understanding of spiritual care, aligning it closely with SCPs' lived practices and realities of care in diverse contexts.

\newdef

\subsection{SPIRIT: A Design Framework for Technological Spiritual Care Interventions}\label{sec:SPIRIT} 
We now present a framework that we will refer to as \textbf{SPIRIT}: \textbf{S}piritual \textbf{P}rerequisites and \textbf{I}nnovative dimensions fo\textbf{R} \textbf{I}ntegrating \textbf{T}echnology. SPIRIT can support the creation of helpful digital tools and interventions for spiritual care by supporting designers to: identify which \textit{phase of care} to target; evaluate \textit{contextual prerequisites} to care; and consider \textit{how} the system's intended design goals and components relate to our definition and design dimensions.

\subsubsection{Phases of Spiritual Care}
\citeauthor{bezabih_meeting_2025} outline a need for both spiritual care practice and technology to support a three-phased model of care~\cite{bezabih_meeting_2025}, including: 

\begin{enumerate}
    \item \textbf{Spiritual needs assessment:} First, \textit{screening} for spiritual distress to identify patients who may benefit from spiritual care~\cite{labuschagne_testing_2025,king_determining_2017,fitchett_developing_2025}, and next \textit{needs assessment} of the specific spiritual experiences, issues, crises, or pain points faced by clients~\cite{anandarajah_spirituality_2001,borneman_evaluation_2010,shields_spiritual_2015,kestenbaum_spiritual_2022,fitchett_development_2019,labuschagne_development_2024}.
    \item \textbf{Spiritual intervention:} Applying established spiritual care methods, techniques, and goals, e.g., ``being with'', gentle touch, or deep listening~\cite{massey_what_2015}.
    \item \textbf{Outcomes assessment:} Understanding whether or not selected technique(s) were effective at addressing identified need(s) and following up if continued care is required~\cite{ahluwalia_macra_2022}.
\end{enumerate}

As an adaptation of language used in \citeauthor{slovak_hci_2024}'s framework for digital mental health systems (see Sec.~\ref{sec:digitization}, we will treat each of the three phases of spiritual care above as components (level 2) which can consist of one or more technical capabilities (level 1) within a complete intervention system (level 3). A complete system will require all three components, whereas an implementation (level 4) can be completed only within a specific, situated context (such as a specific hospital, prison, military unit, school, homeless shelter, faith community, etc.). Designers should begin by identifying which component they should target.

\subsubsection{Contextual Prerequisites}
Spiritual care is not a transaction of information or support, but rather relational accompaniment and care that depends upon pre-conditions. While this is true in offline, digital, and hybrid contexts, we will focus this discussion around digital contexts for purposes of illustration. Before any specific design feature can be effective, our results suggest that the sociotechnical environment must satisfy three foundational prerequisites: the cultivation of safe space, a person's openness to care, and the discernment and articulation of spiritual needs.

\paragraph{Safe Spaces} The concept of safe spaces is not new in HCI. For example, one common case explored in research pertains to marginalized users interacting online. \citeauthor{squires_rethinking_2002} proposes that racially marginalized users often cannot participate safely in the public sphere online, but can find more safety in ``marginal'' publics (enclaves, counterpublics, or satellite publics) that offer protection from hegemonic powers that dominate mainstream spaces~\cite{squires_rethinking_2002,smith_governance_2024}. Prior work that focuses on safety for female~\cite{naseem_designing_2020}, LGBTQ+~\cite{scheuerman_safe_2018,dym_social_2020}, or religious minority~\cite{mahmood_multiple_2025} users has also conceptualized safe spaces as precarious sociotechnical environments where safety is not an inherent platform feature, but rather an active, community-led achievement maintained through the alignment of privacy-enhancing designs and strict social norms to protect marginalized identities from both platform-level vulnerabilities and broader societal harms.

Patients dealing with serious and life-threatening physical or mental illness, or persons experiencing spiritual crises, are often vulnerable and stigmatized. To consider the spiritual needs of vulnerable patients in palliative care, for example, \citeauthor{ahmadpour_how_2023} have also argued that the ``efficiency'' characteristic of healthcare systems must be de-prioritized in favor of slower safety and trust-building design choices~\cite{ahmadpour_how_2023}. Taken together, this body of prior work aligns with our participants' concerns about general-purpose social media having ``negative'' energy that may not be conducive to spiritual care---especially for marginalized, stigmatized, or vulnerable users. Instead, safe spaces, whether offline or online, likely need to be protected enclaves that are socially constructed and upheld by caring, nonjudgmental users; this can set the stage so that people seeking care can expose their ``inner self'' without fear of judgment and sustain meaningful spiritual connection to themselves, others, and/or what they believe to be sacred.

\paragraph{Openness to Care} However, the existence of a safe environment does not guarantee engagement; the user must also possess \textit{openness to care}. As detailed in our findings (Sec.~\ref{sec:prereqs}), openness is not a static trait but a fluctuating state. Internal barriers—such as a desire to control uncontrollable outcomes, the denial of mortality, or rigid theological frameworks (e.g., interpreting suffering solely as divine punishment)—can function as ``maladaptive spiritual responses'' that blind individuals to their own need for support. For designers, this implies that the \textit{timing} of an intervention is as critical as its content~\cite{nahum-shani_just--time_2018,amershi_guidelines_2019}. A system that attempts to intervene when a user is in a state of high resistance or denial may fail or even cause harm~\cite{nahum-shani_just--time_2018}. Therefore, technologies must be sensitive to the user's readiness, potentially employing non-intrusive invitations rather than prescriptive solutions.

\paragraph{Discernment and Articulation of Needs.} Finally, spiritual distress is often experienced as ineffable; users may feel a profound sense of angst, grief, or disconnection but lack the vocabulary to identify it as a spiritual need~\cite{voetmann_verbalizing_2022,puchalski_improving_2009}, or even the ability to articulate a need for support~\cite{smith_what_2021}. Unlike physical symptoms which can often be ``selected from a menu,'' spiritual needs require deep introspection to identify---often with the professional support of a skilled spiritual care generalist (e.g., for spiritual distress screening~\cite{fitchett_developing_2025}) or specialist (e.g., for in-depth needs assessment and intervention~\cite{anandarajah_spirituality_2001,borneman_evaluation_2010}). If a technological interface demands precise input to function (e.g., a search query or a specific help request), it may fail users who are in the midst of the ``fog'' of crisis. This suggests a need for reflective design components that scaffold the discernment process~\cite{baumer_reflective_2015,boehner_interfaces_2008} helping users bridge the gap between vague sensations of distress and actionable requests for care.

\paragraph{Design Questions to Support Prerequisites} As a second step in applying SPIRIT, we offer the following questions for designers to consider before creating new tools, products, or systems:

\begin{itemize}
\item \textbf{Safe Space:} How can spiritual care intervention tools or systems cultivate or become embedded within a perceived safe space for clients?
\item \textbf{Openness to Care:} When and how are clients most open to care? Are there contextual factors that influence their openness, adaptability, or willingness to engage with spiritual care that be addressed through either non-technical or technical means?
\item \textbf{Discernment and articulation of needs:} How does the intervention ensure that stakeholders are able to get in touch with their own genuine spiritual needs---rather than misidentification or imposition of a spiritual need, or a focus on other physical or mental health needs (which may require referrals to other more fitting types of care)?
\end{itemize}

\subsubsection{Design Dimensions for Digital Spiritual Care}
Our results in Sec.~\ref{sec:designdimensions} also provide six design dimensions that are crucial for effective technologically-facilitated delivery of spiritual care: appropriate degree of technology use, loving presence, meaning-making, location, degree of relational closeness, and temporality.

\paragraph{Appropriate Degree of Technology Use} 
\citeauthor{baumer_when_2011} published a provocative paper in 2011, suggesting that the implication to \textit{not} design technology is just as important as implications for how \textit{to} design technology~\cite{baumer_when_2011}. For example, in Sec.~\ref{sec:appropriatedegree}, H4 said, \textit{``God help us if it's just robots taking care of people at end of life,''} indicating profound disapproval for the use of AI or robotics to provide care during the tender, vulnerable moments leading up to death. This opinion connects to a major contemporary debate for HCI regarding the degree to which Artificial Intelligence (AI) can be used to supplement \textit{v.s.} replace healthcare professionals. While some work has begun to explore applications of AI for spiritual practices like prayer or meditation~\cite{kwon_spiritual_2024,nguyen_ai-driven_2024}, recent scholarly work such as~\cite{moore_expressing_2025,smith_designing_2024} critiques the replacement of human mental health or spiritual care workers due to the fact that AI generates inappropriate and stigmatizing outputs, cannot provide human connection, and can trigger psychosis~\cite{fieldhouse_can_2025} or in the worst case, suicide~\cite{roose_can_2024,yang_family_2025}. Although the methods and results of the present paper are inadequate to prescribe \textit{exactly} how and when to integrate technology (including AI) for spiritual care, they do invite designers to carefully consider the degree to which it is or is not appropriate to do so, especially considering ethical commitments of the profession to provide whole-person centered care.

\paragraph{Loving Presence and Meaning-Making}
Our results also underscore that spiritual care is fundamentally an act of meaning-making rooted in loving presence and relational accompaniment. We interpret this through \citeauthor{mekler_framework_2019}'s framework of meaning in HCI, particularly the dimensions of \textit{resonance} and \textit{connectedness}~\cite{mekler_framework_2019}. When our participants described the ``energetic quality'' of a calm, grounded, non-judgmental provider, they were effectively describing an interaction that fosters deep \textit{resonance}---an immediate, felt sense of significance---and \textit{coherence}, allowing care receivers to make sense of their experience without the need to be ``fixed.'' While this quality of loving presence may lie beyond what technology can simulate, design choices nonetheless play a critical role in scaffolding the human effort required to convey it. For instance, in contrast to trends toward frictionless automation, recent work argues that \textit{significance} arises from effortful communication. This includes digital affordances for crafting personalized animations~\cite{zhang_auggie_2022}, investing time to write thoughtful messages rather than using rapid reactions~\cite{smith_thoughts_2023}, or visual metaphors that represent the act of simple presence~\cite{kaur_sway_2021}.

\paragraph{Location, Relationship, and Temporality}
Finally, the sociotechnical context of care is defined by the interplay of \textit{location}, \textit{relational closeness}, and \textit{temporality}. While traditional spiritual care is often located at the physical bedside, our findings highlight how spiritual care stakeholders feel that technology can expand this ``location'' into hybrid or digital spaces, allowing care to traverse physical distances. Technological expansion of care can likewise alter its \textit{temporality}, decoupling it from strictly synchronous encounters. Participants noted how asynchronous digital artifacts like texts or recorded legacies can allow support to carry \textit{``ideas through space and time''} (P9) or help a person's spirit \textit{``last beyond me''} (P3), akin to the enduring nature of sacred texts. However, these affordances must be calibrated against the \textit{degree of relational closeness}~\cite{smith_i_2020}. Designers must determine if a system is intended for high-intimacy strong ties~\cite{krackhardt_strength_1992} (e.g., a close dyadic bond between partners) or the lower-intimacy, potentially asynchronous support of weak ties~\cite{granovetter_strength_1973} in a broader network (e.g., a distant high school friend or kind stranger online). Ultimately, an effective design should match the interaction mode (synchronous/asynchronous) and setting (remote/co-located) to the specific relational depth required by the spiritual need.

\paragraph{Design Questions to Support Innovative Dimensions} As a second step in applying SPIRIT, we offer the following questions for designers to consider during design processes (which are closely tied to Figure~\ref{fig:dimensions}):

\begin{itemize}
    \item \textbf{Technology}: To what extent can and should spiritual care be translated through technology design for a given component and its capabilities? When and what type of technology is appropriate? 
    \item \textbf{Presence}: How can the relational aspect of spiritual care be better supported through technology? Or what part(s) of social companionship could be translated (or not) into design experiences? 
    \item \textbf{Meaning-making}: How can spiritual care components or capabilities help clients to care for their \textit{spirit} or \textit{inner self} through facilitating reflective experiences, resources for finding purpose and inner strength, and making sense of their spiritual struggles?
    \item \textbf{Location}: Which spiritual care components and capabilities can be best supported in-person and which virtually? When is remote or in-person care necessary, and why? 
    \item \textbf{Relationship}: How can spiritual care components and capabilities facilitate the relational aspect of spirituality with others--and \textit{which} others (e.g., known family, friends, and caregivers, or strangers)?
    \item \textbf{Temporality}: How can different components and capabilities facilitate care in the moment, v.s. long-range future or legacy considerations? 
\end{itemize}

\subsection{Hypothetical Application of the SPIRIT Framework}\label{sec:example} To demonstrate the practical utility of SPIRIT, we will apply it to a critical gap in current clinical practice: the lack of effective screening tools for identifying patients in spiritual crisis~\cite{glauser_machine_2017}. Without systematic identification, patients (especially in outpatient care settings) can suffer without needed referrals to professional care~\cite{king_determining_2017,labuschagne_testing_2025}. We will illustrate the hypothetical development of a \textit{digital spiritual distress screening tool} to illustrate how SPIRIT could guide design decisions.

\paragraph{Step 1: Identify Phase} The first step is to situate the technology within the care continuum. This hypothetical tool targets the \textbf{spiritual needs assessment} phase, and even more specifically, patient \textbf{screening}.\footnote{\textbf{Screening} is a lightweight process that is usually completed by clinicians who do \textit{not} specialize in spiritual care---or in some cases, paper or digital survey forms~\cite{fitchett_developing_2025}. \textbf{Needs assessment}, on the other hand, is performed by trained specialists, and is an in-depth, relational activity that leads to the selection and development of targeted interventions~\cite{shields_spiritual_2015,kestenbaum_spiritual_2022,fitchett_development_2019,labuschagne_development_2024}.} It could operate as a level 2 component composed of specific level 1 capabilities, such as Natural Language Processing (NLP) to analyze patient portal messages for linguistic markers of distress, or active questions or journaling prompts that screen for existential angst. The goal of this component should not be to resolve the crisis (which would be the \textit{intervention} phase), but to accurately flag it for human review.

\paragraph{Step 2: Evaluate Contextual Prerequisites} Before implementation, the design must account for the sociotechnical pre-conditions of care. 

\begin{itemize} 
\item \textbf{Openness:} Because openness is a fluctuating state, the system cannot simply extract data; it must invite engagement. The interface might employ ``just-in-time'' logic~\cite{nahum-shani_just--time_2018} to offer screening prompts only when the patient exhibits behavioral signals of readiness, avoiding the harm of forcing spiritual confrontation during acute denial. 
\item \textbf{Safe Space:} To cultivate safety, the tool could explicitly decouple spiritual data from the general medical record, assuring patients that sensitive disclosures about "loss of meaning" or "anger at God" are confidential and viewed only by spiritual care specialists, protecting them from perceived judgment by medical staff or family caregivers who view their data. 
\item \textbf{Discernment:} Recognizing that spiritual pain is often ineffable~\cite{voetmann_verbalizing_2022}, the UI should avoid rigid search queries in favor of scaffolded reflection. Capabilities might include visual mood boards or metaphorical prompts that help users bridge the gap between vague sensations of distress and an articulable request for support. \end{itemize}

\paragraph{Step 3: Apply Design Dimensions} Finally, designers must calibrate the tool using the six design dimensions. For a screening tool, the \textbf{appropriate degree of technology} is paramount: the system must function strictly as a triage mechanism, not a replacement for human care. (While \textbf{presence} and \textbf{meaning-making} are crucial during intervention itself, they are less relevant during screening.) It should identify distress but avoid generating algorithmic advice, which risks offering stigmatizing~\cite{moore_expressing_2025} or theologically hollow platitudes. \textbf{Temporality} is also critical; the system must differentiate between urgent signals requiring synchronous alerts to a chaplain (e.g., despair at end-of-life) versus chronic spiritual struggles suitable for asynchronous logging. Finally, \textbf{relational closeness} could dictate a helpful referral pathway: high-acuity alerts should route to professional SCPs (weak ties with expertise), whereas lower-level social needs might prompt the user to connect with their own faith community or family (strong ties).

We note briefly that the example above is only one initial possibility for a needs assessment component (for instance, other possibilities might include tools that assist clinicians with verbal spiritual needs assessment in real time during appointments or ambient sensing tools deployed in home environments or mobile devices). A similar approach can guide the design of components for interventions or outcomes assessment---all of which are needed in a complete intervention system for spiritual care. Intervention systems (level 3) and future implementations (level 4) should likewise be studied and pursued in collaboration with professional spiritual care providers to best align HCI and clinical work. 

\subsection{Limitations and Future Work}\label{sec:futurework}
\paragraph{Sample Limitations} Although SPIRIT offers a structured approach for designing spiritual care interventions, there are several limitations of our methods. Across both samples, this study represents the voices of $n=56$ participants. Although this is relatively sizable for a qualitative study, and participants were of diverse R/S backgrounds, we cannot claim it is necessarily representative of all spiritual care stakeholders. In particular, while this study includes perspectives of multiple stakeholder groups, member checking was performed only with professional SCPs who work across a variety of healthcare and community settings, ranging from clinical contexts (e.g., chaplains in palliative care or intensive care units), private practice (e.g., spiritual directors working 1:1 with clients), or communities (e.g., homelessness shelters, R/S faith leaders, educational institutions.  We acknowledge that we did not have the opportunity to work with SCPs working in military or prison settings, for example. The delicate contextual considerations of these vulnerable settings warrant future in-depth qualitative work in collaboration with organizations like the US Department of Veteran's Affairs (VA) or Department of Defense branches, or within specific prison settings. Future work can also confirm the insights presented in this work with quantitative methods and larger sample sizes across different types of stakeholders.

\paragraph{Validating the Framework} Moreover, future work should validate the framework and its efficacy for supporting design processes by using the framework to support new collaborative projects with clinicians. Although interdisciplinary clinical work raises distinct challenges, a ``team science'' approach can improve real world implementation of HCI advances and improve broader impacts in society~\cite{agapie_using_2022,lyon_bridging_2023,agapie_conducting_2024}. We therefore advocate for future work to continue involving SCPs and other stakeholders directly in the design and implementation of future digital spiritual care systems. 

\paragraph{A Need for Better Outcomes Assessment} Apart from emerging tools such as the ``Feeling Heard and Understood'' scale~\cite{ahluwalia_macra_2022}, the field currently lacks robust methodologies for concretely assessing outcomes of spiritual care and of the impact of technology on spiritual well-being. Future research in both spiritual care and HCI should develop metrics and techniques for this purpose so that impacts of technological and non-technological spiritual care interventions can be better understood through scientific methods. 

\paragraph{Connecting with Communities of Practice} Finally, exploring the harmonious integration of traditional spiritual tools with modern technology, and ensuring accessibility for individuals with diverse needs, cultural backgrounds, illnesses, or  disabilities will enhance the inclusivity and effectiveness of technospiritual designs. Addressing these gaps will advance the field and improve the development of technologies for supporting spiritual health. Developers or HCI practitioners interested in this domain can connect with communities of practice such as the \href{https://spiritedhci.org/}{SPIRITED Collective} (an HCI-led effort to tackle design challenges at the intersection of interactive technology and R/S) or the  \href{https://www.spiritualinnovation.org/}{Spiritual Innovation Directory} by the Sacred Design Lab (a non-profit design studio), or with organizations such as \href{https://telechaplaincy.io/}{Telechaplaincy.io}, \href{https://transforming-chaplaincy.mn.co/}{Transforming Chaplaincy}, and the \href{https://chaplaincyinnovation.org/}{Chaplaincy Innovation Lab}, which all focus on supporting the evolution of professional spiritual care practice to meet modern human needs given modern contexts.

\section{Conclusion}
The professional field of spiritual care has only recently begun adopting digital technologies to improve access to care and expand models of delivery. This paper aims to support these efforts and to better leverage HCI methods and expertise toward effective integration and implementation of technology. We re-analyzed data used to derive a definition of ``spiritual support'' for HCI in~\cite{smith_what_2021}, and collected new interviews with spiritual care providers in order to contribute: (1) an updated definition of ``spiritual care''; and (2) the SPIRIT framework which provides a cohesive structure for designing digital tools and interventions for spiritual care. By deriving a framework inductively from stakeholder data, this contribution realizes the promise of Grounded Theory Method~\cite{muller_curiosity_2014} to generate foundational insights that can guide design in the emerging domain of digital spiritual care.

\begin{acks}
We sincerely thank our anonymous reviewers who provided insightful comments on earlier drafts of this paper and all participants who have contributed data to both parts of this study. We are also appreciative to Loren Terveen, Susan O'Conner-Von, Anne-Marie Snider, and George Handzo who contributed to  data collection, analysis, and feedback on this work at various stages. The initial work on the Spiritual Support definition~\cite{smith_what_2021} was funded by CaringBridge.org. The new interviews presented in this paper were funded by the John Templeton Foundation, grant \href{https://www.templeton.org/grant/expanding-models-of-delivery-for-online-spiritual-care}{\#62930}.
\end{acks}

\bibliographystyle{ACM-Reference-Format}
\bibliography{JTF2}
\end{document}